\documentclass[11pt]{article}%
\usepackage{amsfonts}
\usepackage{amsmath}
\usepackage{amsthm}
\usepackage{amssymb}
\usepackage{graphicx}

\newtheorem{theorem}{Theorem}

 \newtheorem{lemma}[theorem]{Lemma}

\newtheorem{proposition}[theorem]{Proposition}

\def\N{\mathbb{N}}
\def\ZZ{\mathbb{Z}}

\def\cG{\mathcal G}

\def\C{{\mathbb C}}

\def\Q{{\mathbb Q}}
\def\Z{{\mathbb Z}}
\def\R{{\mathbb R}}

\def\GG{{\mathfrak G}}

\def\mand{\qquad \mbox{and} \qquad}

\def\diam{\mathrm{diam}\, }

\def\\{\cr}
\def\({\left(}
\def\){\right)}

\def\<{\langle}
\def\>{\rangle}

\def\le{\leqslant}
\def\ge{\geqslant}

\begin{document}

\title{\sc Parameters of Integral Circulant Graphs and Periodic
Quantum Dynamics}

\author{
{\sc Nitin Saxena}\\
{Centrum voor Wiskunde en Informatica}\\
{P.O. Box 94079, 1090 GB Amsterdam, The Netherlands}\\
{\tt ns@cwi.nl}\\
\and
{\sc Simone Severini} \\
{Institute for Quantum Computing}\\
{University of Waterloo, N2L 3G1 Waterloo, Canada}\\
{\tt simoseve@gmail.com} \\
\and
{\sc Igor E.~Shparlinski} \\
{Department of Computing, Macquarie University} \\
{Sydney, NSW 2109, Australia} \\
{\tt igor@ics.mq.edu.au}}

\maketitle
\begin{abstract}

The intention of the paper is to move a step towards a
classification of network topologies that exhibit periodic quantum
dynamics. We show that the evolution of a quantum system, whose
hamiltonian is identical to the adjacency matrix of a circulant
graph, is periodic if and only if all eigenvalues of the graph are
integers (that is, the graph is \emph{integral}). Motivated by this
observation, we focus on relevant properties of integral circulant
graphs. Specifically, we bound the number of vertices of integral
circulant graphs in terms of their degree, characterize
bipartiteness and give exact bounds for their diameter.
Additionally, we prove that circulant graphs with odd order do not
allow perfect state transfer.
\end{abstract}

\section{Introduction}

Circulant graphs have a vast number of uses and applications to
telecommunication network, VLSI design, parallel and distributed
computing (see~\cite{Hw} and the references therein).

A graph is integral if all the eigenvalues of its adjacency matrix
are integers (see~\cite{BCRASS} for a survey on integral graphs).

Here, we first show that the evolution of a quantum system, whose
hamiltonian is identical to the adjacency matrix of a circulant
graph, is periodic if and only if the graph is integral. Then,
motivated by this observation, we focus on relevant properties of
integral circulant graphs.

The intention of the paper is to move a step towards a
classification of network topologies which exhibit periodic quantum
dynamics. For certain quantum spin systems with fixed
nearest-neighbour couplings, periodicity is a necessary condition
for \emph{perfect state transfer}, that is, for transferring a
quantum state between sites of the system, with the use of a
\emph{free} evolution and without dissipating the information
content of the state (see~\cite{CDEL}, for more information on this
topic).

The main mathematical results of the paper are the following:
\begin{itemize}
    \item we bound the order of connected integral circulant graphs as a
function of the degree (Section~\ref{sec:deg_ord}, Theorem~\ref{thm:Nk});
    \item we
characterize bipartite integral circulant graphs (Section~\ref{sec:bipart},
Theorem~\ref{thm:Nkb});
    \item we prove tight lower and upper bounds on the
diameter of integral circulant graphs (Section~\ref{sec:diam}, Theorem
\ref{thm-diam-bounds} and Theorem \ref{thm-diam-bounds-tight}).
\end{itemize}

Given the properties of circulant graphs, the proofs are based on
elementary number theory. In the last section we show that circulant
graphs with odd order do not allow perfect state transfer. However,
we do not have a characterization of integral circulant graphs
allowing perfect state transfer. This is left as an open problem.

\section{Background on Circulant Graphs}
\label{sec:background}

A \emph{graph} $\cG=(V(\cG),E(\cG))$ is a pair whose elements are
two sets, $V(\cG)=\{1,2,\ldots,n\}$ and $E(\cG)\subset V(\cG)\times
V(G)$. The elements of $V(\cG)$ and $E(\cG)$ are called
\emph{vertices} and \emph{edges}, respectively. We assume that
$\{i,i\}\notin E(\cG)$ for all $i\in V(\cG)$. Two vertices $i,j$ of
a graph are said to be \emph{adjacent} if $\{i,j\}$ is an edge; the
edge $\{i,j\}$ is then \emph{incident} with the vertices $i,j$.

The \emph{adjacency matrix} of a graph $\cG$ is the matrix $A(\cG)$
such that $A(\cG)_{i,j}=1$ if $\{i,j\}\in E(\cG)$ and
$A(\cG)_{i,j}=0$ if $\{i,j\}\notin E(\cG)$. The \emph{spectrum} of a
graph $\cG$ is the collection of eigenvalues of $A(\cG)$, or
equivalently, the collection of zeros of the characteristic
polynomial of $A(\cG)$, see~\cite{CDS}. We denote by
sp$(\cG)=(\lambda_{0}(\cG),\ldots ,\lambda_{n-1}(\cG))$ the spectrum
of a graph $\cG$ in the non-increasing (wrt modulus) ordering. We simply write
$\lambda_{0},\ldots,\lambda_{n-1}$ when $\cG$ is clear from the
context.

Let $S=\{s_{1},s_{2},\ldots,s_{k}\}$ be a set of $k$
integers in the range
$$
1\leq s_{1},s_{2},\ldots,s_{k}<n.
$$
Since
we consider only \emph{undirected graphs}, we assume that
$s\in S$ if
and only if $n-s\in S$.

A \emph{circulant graph} $\cG =G(n;S)$ is a graph on the set of $n$
vertices $V(\cG)=\{v_{1},\ldots,v_{n}\}$ with an edge incident with
$v_{i}$ and $v_{j}$ whenever $|i-j|\in S$, see~\cite{Hw}. The set
$S$ is said to be the \emph{symbol} of $\cG$. In particular, $k=\#S$
is the \emph{degree} of a circulant graph $G(n;S)$.

Let $\Z_n$ denote the \emph{residue ring} modulo $n$
and
let $\Z_n^\ast$ be the  multiplicative group of $\Z_n$.

Notice that a circulant graph $G(n;S)$ is a {\it Cayley graph\/} of the
additive group of $\Z_n$ with respect to the {\it Cayley set\/} $S$.

We recall that a Cayley graph with respect to a finite group $\GG$
and a set $S \subseteq \GG$, such that it contains $-w$ for every $w
\in S$, is a graph on $n = \# \GG$ vertices, labeled by elements of
$\GG$ where the vertices $u$ and $v$ are connected if and only if
$u-v \in S$ (or equivalently, $v - u \in S$).

A \emph{path} in a graph is a finite sequence of vertices which are
connected by an edge. A \emph{connected graph} is a graph such that
there is a path between all pairs of vertices. It is easy to show that a
circulant graph $G(n;S)$ with symbol $S=\{s_{1},s_{2},\ldots,s_{k}\}$
is connected if and only if $\gcd(n,s_{1},s_{2},\ldots,s_{k})=1$.

The adjacency matrix of a circulant graph is diagonalized by the
Fourier transform at the irreducible representations over the group
$\Z_n^\ast$ (which is a Vandermonde matrix). Lemma~\ref{lem:Eigen}
is based on this observation.

Let $\omega_{n}=\exp\(  2\pi\iota/n\)$ where $\iota=\sqrt{-1}$.

\begin{lemma}
\label{lem:Eigen}The spectrum of a circulant graph $\cG=G(n;S)$ on
$n $ vertices with symbol $S$ is
\begin{equation}\label{eq-lambda-in-terms-of-omega}
\lambda_{j}=\sum_{s\in S}\omega_{n}^{js},
\end{equation}
where $0\leq j\leq n-1$.
\end{lemma}

By  Lemma~\ref{lem:Eigen}, the eigenvalues of a circulant graph are
just the sum over $S$
of the irreducible characters of $\Z_n^\ast$. The eigenvectors
are also easily available. In fact, it is straightforward to see that the
eigenvector corresponding to the eigenvalue $\lambda_{j}$ has the form
$v_{j}=[1,\omega^{j},\ldots,\omega^{j(n-1)}]^{T}$.

\section{Integral Circulant Graphs and Periodic Quantum Dynamics}
\label{sec:PerQuantChan}

A {\em quantum spin system} associated to a graph $\cG$ can be
defined by attaching a spin-$\frac{1}{2}$ particle to each of the
$n$ vertices of $\cG$. The Hilbert space assigned to the system is
then $\mathcal{H}\cong(\C^{2})^{\otimes n}$.

This system can be interpreted as a noiseless quantum channel, whose
Hamiltonian is identical to the adjacency matrix of $\cG$ itself.
 From another perspective, its evolution can be seen as a
continuous-time quantum walk on $\cG$. Some properties of such
dynamics on circulant graphs have been studied in~\cite{am}.

As observed in~\cite{CDEL}, the dynamics of the system is {\em
periodic} if for every state $|\psi\rangle\in\mathcal{H}$, there
exists $p\in\R$, $0<p<\infty$, for which $|\langle\psi|e^{-\iota
A(\cG)p}|\psi\rangle|=1$. The number $p$ is the {\em period} of the
system.

In general, assuming that the initial state was
$|\psi(0)\rangle=\sum_j\alpha_j|\lambda_{j}\rangle$, we can express
as follows the state of the system at generic time $t$:
\[
|\psi(t)\rangle=e^{-\iota H_G
t}|\psi(0)\rangle=\sum_{j}\alpha_{j}e^{-\iota
t\lambda_{j}}|\lambda_{j}
\rangle
\]
where $|\lambda_{j}\rangle$ is an eigenvector of $A(\cG)$ with
eigenvalue $\lambda_{j}$ and $\alpha_{j}\in\C$.
Thus, the periodicity condition
$|\psi(t)\rangle=e^{-\iota\phi}|\psi(0)\rangle$ ($\phi$ is a phase)
gives us that  for every $\lambda_j\in\text{sp}(G)$ we have:
\[
\lambda_{j}t-\phi=2\pi r_{j}, \text{~~ for some }r_{j}\in\ZZ.
\]
Therefore, for every quadruple
$\lambda_{i},\lambda_{j},\lambda_{r},\lambda_{s}\in$ sp$(G)$ (with
$\lambda_{r}\neq\lambda_{s}$), it
follows that
\begin{equation}
\label{eq:lambdas}
\frac{\lambda_{i}-\lambda_{j}}{\lambda_{r}-\lambda_{s}}\in\Q.
\end{equation}

We now show that~\eqref{eq:lambdas} implies the integrality of the
underlying graph.

\begin{theorem}
\label{thm:periodic} Let $\cG=G(n;S)$ be a circulant graph on $n\ge 4$
vertices with symbol $S$. If $\cG$ has at least four distinct
eigenvalues and all of them
satisfy the condition~\eqref{eq:lambdas} then $\cG$ is integral.
\end{theorem}

\begin{proof}
Let $k  = \# S$ be the degree of $\cG$.
By Lemma~\ref{lem:Eigen}, $\lambda
_0=k$.  It is clear then that $\lambda_1,\ldots,\lambda_{n-1}$ are
all different from $\lambda_0$. If sp$(\cG)$
satisfies~\eqref{eq:lambdas}, then for all $i\in\{1,\ldots,n-1\}$,
we have $$
\frac{\lambda_i-k}{\lambda_1-k}\in\Q.
$$
Therefore, $\lambda_i=a_i\lambda_1+b_i$ for some $a_i, b_i\in\Q$.

We now show that   $\lambda_1\in\Q$. For this we consider three cases:

{\bf Case 1:} Suppose $n = p$, a prime.

Then the minimal polynomial of $\omega_n$ over $\Q$ is $1+X+\ldots+X^{n-1}$.
Since $\cG$ has at least four distinct eigenvalues we can find $2\le
j<h\le(n-1)$ such
that $\lambda_0$, $\lambda_1$, $\lambda_{j}$ and $\lambda_{h}$ are
all distinct.

Suppose that $\lambda_1\not\in\Q$.
{From}~\eqref{eq:lambdas} we have that
$\lambda_{j}=a_{j}\lambda_1+b_{j}$ for some $a_{j}, b_{j}\in\Q$.
Applying~\eqref{eq-lambda-in-terms-of-omega} we get
$$
\sum_{s\in S}\omega_n^{js}=a_{j}\sum_{s\in S}\omega_n^s+b_{j}.
$$
In the last identity we can replace each exponent $js$ with its
smallest positive
residue $r_{j,s}$ modulo $n$, which in turn   means the following
divisibility of polynomials
$$
1+X+\ldots+X^{n-1}~\Big|~\sum_{s\in S}X^{r_{j,s}}-a_{j}\sum_{s\in S}X^s-b_{j}.
$$
Since the nonzero polynomial on the right hand side is of degree at most $n-1$
and since $\lambda_1\not=\lambda_{j}$, we obtain
$$
1+X+\ldots+X^{n-1} = \sum_{s\in S}X^{r_{j,s}}-a_{j}\sum_{s\in S}X^s-b_{j} ,
$$
which implies
$-a_{j}=-b_{j}=1$. Thus,
$\lambda_{j}=-\lambda_1-1$. Applying the same argument on $\lambda_{h}$ we get,
$\lambda_{h}=-\lambda_1-1$, thus implying $\lambda_{h}=\lambda_{j}$. This
contradiction shows that $\lambda_1\in\Q$ in this case.

{\bf Case 2:} Suppose $n=p^r$, a power of a prime $p$, where $r\ge 2$.

Now focus on the set of eigenvalues:
$$\left\{\lambda_{p^{r-1}},\lambda_{2p^{r-1}},\ldots,\lambda_{(p-1)p^{r-1}}\right\}$$
Suppose that $\lambda_1\not\in\Q$.
Clearly, $\lambda_{p^{r-1}}$ cannot be rational (otherwise $\lambda_1\in\Q$).
The above eigenvalues can be described as:
$$
\lambda_{ip^{r-1}}= \sum_{s\in S}\omega_{n}^{ip^{r-1}s}=\sum_{s\in
S}\omega_p^{is}.
$$
Thus, this case now reduces to the prime case above and
shows that $\lambda_1$ is rational.

{\bf Case 3:} Suppose $n$ has two distinct prime factors $p, q$.

We have that for all
$i\in\{1,\ldots,n-1\},\ \lambda_i=a_i\lambda_1+b_i$, for some $a_i,
b_i\in\Q$. Thus,
\begin{equation}\label{eq-equal-fields}
\Q(\lambda_1)=\ldots=\Q(\lambda_{n-1}).
\end{equation}
Observe that $\lambda_{n/p}\in\Q(\omega_p)$ and $\lambda_{n/q}\in\Q(\omega_q)$.
But the equation~\eqref{eq-equal-fields} implies that
$\lambda_{n/p}\in\Q(\lambda_{n/q})$.
Thus, $\lambda_{n/p}\in\Q(\omega_p)\cap\Q(\omega_q)$. It can be shown that
$\Q(\omega_p)\cap\Q(\omega_q)=\Q$ since $p, q$ are coprime. Thus,
$\lambda_{n/p}\in\Q$ and then the equation~\eqref{eq-equal-fields}
forces $\lambda_1\in\Q$.

Thus, in all the cases $\lambda_1\in\Q$ and hence all the $n$
eigenvalues are rational. Since they are also algebraic integers,
this further implies the desired result.
\end{proof}

It is plausible that the method of proof of
Theorem~\ref{thm:periodic} can be extended to other classes of
Cayley graphs.

In the light of Theorem~\ref{thm:periodic}, in the next sections we
consider parameter of circulant integral graphs. Before doing that,
we now give a characterization of these graphs, which is due to
So~\cite{So} (and which is naturally based on
Lemma~\ref{lem:Eigen}). This is our main technical tool.

Let \[ G_{n}(d)=\{k~\mid~1\leq k\leq n-1,\gcd(k,n)=d\}, \] be the set of all
integers less than $n$ having same greatest common divisor $d$ with
$n$. In particular $\#G_{n}(d)=\varphi(n/d)$, where, as usual, \[
\varphi(m)=\#\{1\leq s\leq m~\mid~\gcd(s,m)=1\} \] denotes the Euler
totient function of a positive integer $m$ (see, for
example,~\cite{HardyWright}).

Notice that the collection $\{G_{n}(d)~\mid~d\mid n\}$ is a partition
of the set
$\{1,2,\ldots,n-1\}$. Notice that $k\in G_{n}(d)$ if and only if
$n-k\in G_{n}(d)$, since $\gcd(k,n)=\gcd(n-k,n)$.

Let $D_{n}$ be the set of all $\tau(n)-1$ divisors $d\mid n$ with
$d\leq n/2$, where, as usual, $\tau(n)$ is the number of positive
integer divisors of $n$.

\begin{lemma}
\label{lem:Char} A
circulant graph $\cG=G(n;S)$ on $n$ vertices with
symbol $S$ is
integral if and only if
\begin{equation}
S=\bigcup_{d\in D}G_{n}(d)
\label{eq:union}
\end{equation}
for some set of divisors $D\subseteq
D_{n}$.
\end{lemma}

Throughout the paper, the implied constants in the symbols `$O$',
`$\ll$ and `$\gg$' are absolute. We recall that $A\ll B$ and $B \gg
A$ is equivalent to the statement that $A=O(B)$ for positive
functions $A$ and $B$).

\section{Degree and Order}
\label{sec:deg_ord}

In this section we prove an upper bound on the number of vertices of
an integral circulant
graph in terms of its degree.

\begin{theorem}
\label{thm:Nk} There is an absolute constant $c>0$ such that for any $k\geq2
$, the largest  number $N(k)$ of vertices of an integral connected
circulant graph $\cG=G(n;S)$
having degree $k$ is bounded by
\[
N(k)\leq \exp\(c\sqrt{k\log\log(k+2)}\log k\).
\]
\end{theorem}

\begin{proof}
By Lemma~\ref{lem:Char}, we see that $S=\bigcup_{d\in D}G_{n}(d)$, for some
set of divisors $D\subseteq D_{n}$. Therefore
\begin{equation}
\label{eq:k and F}
k=\#S=\sum_{f\in F}\varphi(f).
\end{equation}

Given that $\cG$ is connected, we have $\gcd(\{d~\mid~d\in D\},n)=1$.

Noting that for any two divisor $f,F\mid n$ we have
\[
\gcd(n/f,F)\geq\gcd(F/f,F)\geq F/f,
\]
it is easy to prove by induction on $m$ that for any sequence
$f_{1}
,\ldots,f_{m}$ of divisors of $n$ we
have
\[
\gcd(n/f_{1},\ldots,n/f_{m},n)\geq\frac{n}{f_{1}\ldots
f_{m}}.
\]
Therefore
$$
1    =\gcd(\{d~\mid~d\in D\},n)=\gcd\(\{n/f~\mid~f\in
F\},n\)
    \geq n\prod_{f\in F}f^{-1} $$ which leads us to the bound

\begin{equation} n\leq\prod_{f\in F}f. \label{eq:prod div}
\end{equation}
We now recall the well known bound that for some
absolute constant $C>0$,
\begin{equation}
\varphi(f)\gg\frac{f}{\log\log(f+2)} \label{eq:phi},
\end{equation}
see~\cite[Theorem~328]{HardyWright}. Thus we see
from~\eqref{eq:k and F} that
\[ \frac{f}{\log\log(f+2)}\ll k \]
for every
$f\in F$,
which obviously implies that \[ f\ll k\log\log(k+2). \] Now,
using
this bound together with~\eqref{eq:phi} and~\eqref{eq:k and F} again,

we derive $$ k =\sum_{f\in F}\varphi(f)\gg\sum_{f\in
F}\frac{f}{\log\log(f+2)} \gg\sum_{f\in F}\frac{f}{\log\log(k+2)}.
$$
Thus if we denote $$ \sigma=\sum_{f\in F}f $$ then we have
\begin{equation}
\sigma\ll k\log\log(k+2). \label{eq:bound sigma}
\end{equation}

Let $s=\#F$. Then we deduce from~\eqref{eq:prod div}
that
\begin{equation}
n\leq\(  \sigma/s\)  ^{s}. \label{eq:bound
n}
\end{equation}
Since
\[
\sigma=\sum_{f\in F}f\geq\sum_{j=1}^{s}j=\frac{s(s+1)}{2}  ,
\]
we see
that
\begin{equation}
s\ll\sqrt{\sigma}. \label{eq:bound s}
\end{equation}

Since the function $\(  \sigma/x\)  ^{x}$
monotonically increases for
$1\leq x\leq\sigma/e$, we obtain
from~\eqref{eq:bound sigma}
and~\eqref{eq:bound n}
that
\[
n\leq\exp\(  O\(  \sqrt{\sigma}\log\sigma\)  \)  ,
\]
and
recalling~\eqref{eq:bound sigma}, we conclude the
proof.
\end{proof}

On the basis of the arguments used in the proof of
Theorem~\ref{thm:Nk}, we can construct the following table, in which
we list the maximum order of an integral circulant graph of fixed
degree $k=2,\ldots,11$ (this is the sequence A126857
in~\cite{oeis}).
\[
\begin{tabular}
[c]{c|c}
Degree $k$ & Maximum
order $N(k)$\\\hline
$2,3$ & $6$\\
$4,5$ & $12$\\
$6,7$ &
$30$\\
$8,9$ & $42$\\
$10,11$ &
$120$
\end{tabular}
\]

\section{Bipartiteness}
\label{sec:bipart}

In this section we characterize bipartite integral circulant graphs.

Let us denote by $\mu(m)$ the M\"{o}bius function of a positive
integer $m$:
\[
\mu(m)=\left\{
\begin{tabular}
[c]{ll} $0$, & if $m$
has repeated prime factors;\\
$1$, & if $m=1$;\\
$\(  -1\)  ^{k}$, & if $m$ is a product of $k$ distinct primes.
\end{tabular}
\
\right.
\]
For a fixed $k$, there exists a set $F\subset\N$ such that we
have~\eqref{eq:k and F}.
Writing
\[
n=\mathrm{lcm}\left\{f~\mid~f\in F\right\}
\]
and
\begin{equation}
S=\bigcup_{f\in F} G_n\(\frac{n}{f}\)
\label{ess}
\end{equation}
it is not hard to see that that the above defines an integral
circulant graph $\cG=G\(n;S\)  $.
As discussed in \cite{So}, the eigenvalues of $\cG=G\(n;S\)  $ are then:
for $0\leq j\leq n-1$,
\begin{equation}
\lambda_{j}=\sum_{f\in F}\varphi(f)\cdot
\frac{\mu\(f/\gcd(f,j)\)}{\varphi\(f/\gcd\(  f,j\)\)}.
\label{eigv}
\end{equation}
By~\eqref{eigv}, we can determine which integral circulant graphs are
bipartite.

\begin{theorem}\label{thm:Nkb}
An integral circulant graph $\cG=G(n;S)$ on $n$ vertices with
symbol $S$ is bipartite if and only if $n$ is even and $S=\cup_{f\in
F} G_n\(\frac{n}{f}\)$, where
for some number $\ell_0$, the set $\left\{ 2\ell_0/f~\mid~f\in
F\right\}$ contains only odd integers.
\end{theorem}

\begin{proof}
Having degree $k$, the graph $\cG$ is bipartite if and only if it has
an eigenvalue $\lambda_{\ell}=-k$, see~\cite{CDS}.

Suppose $\cG$ is bipartite. On the basis of~\eqref{ess} and~\eqref{eigv},
\[
\lambda_{\ell}=-k=\sum_{f\in F}\varphi\(  f\)\cdot
\frac{\mu\( f/\gcd(f,\ell)\)}{\varphi\(f/\gcd\(f,\ell\)\)}.
\]
Since~\eqref{eq:k and F}, the above equation can hold only if for
every $f\in F$:
\[
\frac{\mu\( f/\gcd(f,\ell)\)}{\varphi\(f/\gcd\(f,\ell\)\)} =-1.
\]
This implies that
\begin{equation}
\label{emm} \mu\(  \frac{f}{\gcd(f,\ell)}\)  =-1
\mand \varphi\(\frac{f}{\gcd\(  f,\ell\)  }\) =1.
\end{equation}
Whence,
\begin{equation}
\label{eff}
\frac{f}{\gcd\(
f,\ell\)  }\in\{1,2\}.
\end{equation}
So, the equation~\eqref{emm} together with~\eqref{eff} gives:
\[
\frac{f}{\gcd\(  f,\ell\)  }=2.
\]
Implying that for every $f\in F$ the ratio
$ 2\ell/f$ is an odd integer.

Also it follows that $n$ is even as $n=\mathrm{lcm}\left\{f~\mid~f\in
F\right\}$. Thus, the theorem
is true in one direction.

Conversely, suppose that $n$ is even and $2\ell_0/f$ is odd for every $f\in F$.
Consequently, the $\ell_0$-th eigenvalue is:
\begin{align*}
\lambda_{\ell_0}&= \sum_{f\in F}\varphi\( f\)\cdot
\frac{\mu\(f/\gcd(f,\ell_0)\)}{\varphi\(f/\gcd\(f,\ell_0\)\)}\\
& = \sum_{f\in F}\varphi\( f\)\cdot\frac{\mu(2)}{\varphi(2)}
=  \sum_{f\in F}\varphi\( f\)\cdot(-1)=  -k
\end{align*}
Thus, $\cG$ is bipartite and the theorem is proved.
\end{proof}

\section{Diameter}
\label{sec:diam}

In this section we prove tight lower and upper bounds on the
diameter of integral circulant graphs.

The \emph{diameter} of a graph $\cG$, denoted by $\diam \cG$, is the
longest among the shortest paths between any two vertices. If $\cG$
is a circulant graph on $n$ vertices then it is clear that $1\leq
\diam \cG \leq n/2$.

For a given degree $k$, the number of vertices of an integral
circulant graph $\cG$ can be $n=\mathrm{lcm}\{f~\mid~f\in F\}$, where $F$
is such that we have given in equation \eqref{eq:k and F}.

Assuming that the columns (and rows) of the adjacency matrix
$A_{\cG}$ of $\cG$ are labelled from $1,\ldots,n$ then the first row
of $A_\cG$ is:

\[
S=\bigcup_{f\in
F}G_{n}\(  \frac{n}{f}\)  =\bigcup_{f\in F}\left\{
i\ \lvert\ 1\leq
i\leq n,\ \gcd(i,n)=\frac{n}{f}\right\}
\]
A right shift of row $S$ gives the subsequent rows of $A_{\cG}$.

Let $X\subseteq\Z_n$ then, for a positive integer,  we define
$$iX=\underset{i\text{
times}}{\underbrace{X+\ldots+X}} =
\{x_1 +\ldots + x_i~\mid~x_1
,\ldots , x_i \in X\}
$$
(where the elements are added modulo $n$). Note that the vertices in
$\cG$ reachable from the vertex $0$ in $1$ step are exactly the
vertices of $S$; the vertices reachable from the
vertex $0$ in $2$ steps are those of $2S$, and so on so forth.
Similarly, if we define $T=S\cup\left\{ 0\right\}$ then the vertices
reachable from the vertex $0$ in $i$ or smaller steps are $ iT$.
Thus, we have:

\begin{lemma}
\label{lem-diam-T-i-times} The diameter of the
circulant graph
$\cG=G(n;S)$ is the least
index $i$ such that $
iT=\Z_n$.
\end{lemma}

\begin{theorem}
\label{thm-diam-bounds} Let $D$ be a set of divisors of $n$ such that
$\gcd(D,n)=1$
and let $t$ be the
size of the smallest set of additive
generators of $\Z_n$ contained
in $D$. Then, for the circulant graph
$\cG=G(n;S)$, where  $S=\cup_{d\in D}G_{n}(d)$,  we have
\[
t\leq\diam  \cG\leq 2t+1.
\]
\end{theorem}

\begin{proof} It is very simple to
show the lower bound. By the hypothesis, it is easy to see that $t$
is the size of the smallest set of generators of $\Z_n$ contained in
$T$. Thus, by Lemma~\ref{lem-diam-T-i-times}, we deduce that $\diam
\cG\geq t$.

We now turn to the upper bound. Let $d_{1},\ldots,d_{t}\in D$ be the
additive generators of $\Z_n$. Without loss of generality, we can
assume that $d_{1}$ is odd. Clearly, $\gcd(d_{1},\ldots,d_{t},n)=1$.
We intend to show that given any $\ell\in\Z_n$ there exist
$x_{0},x_{1},\ldots,x_{2t} \in(\Z_n)^{\ast}$ such that either
$$ d_{1}x_{0}+d_{1}(x_{1}+x_{t+1})+\ldots+d_{t}(x_{t}+x_{2t}) \equiv
\ell \pmod n $$ or \begin{equation} \label{eqn-coprime-solns-2}
d_{1}(x_{1}+x_{t+1})+\ldots+d_{t}(x_{t}+x_{2t}) \equiv \ell \pmod n
.
\end{equation}
Note that this would mean that $(2t+1)T=\Z_n$.
We
now solve one of the above congruences  modulo prime factors of
$n$ and then
``lift'' that solution modulo $n$.

If $2|n$ then we can
put
\[
x_{0} \equiv  x_{1}  \equiv  \ldots  \equiv  x_{2t}  \equiv 1
\pmod 2
\]
and then, depending on the parity of $\ell$ one of the above
equations, say~\eqref{eqn-coprime-solns-2},
   holds modulo $2$.
Suppose $\alpha_{2}$ is the largest index of $2$ dividing $n$. Then
this solution can be Hensel lifted~\cite{LN} to a solution
$(x_{0},\ldots,x_{2t})$ modulo $2^{\alpha_{2}}$.

Next, let $p$ be an odd prime dividing $n$. Since $\gcd(d_{1},\ldots
,d_{t},n)=1$, without loss of generality, we can assume that $p\nmid
d_{1}$. Now we substitute \[ x_{2} \equiv \ldots \equiv x_{t} \equiv
1 \equiv -x_{2+t} \equiv \ldots \equiv -x_{2t} \pmod p \] and
then~\eqref{eqn-coprime-solns-2} simply becomes \[d_{1}(x_{1}
+x_{t+1}) \equiv \ell \pmod p\] or \[ x_{1}+x_{t+1} \equiv \ell\cdot
d_{1}^{-1} \pmod p \] and we can easily find nonzero values of
$x_{1}$ and $x_{t+1}$ modulo $p$. So we have a solution of the
equation~\eqref{eqn-coprime-solns-2} modulo $p$ and it can be Hensel
lifted to a solution modulo $p^{\alpha_{p}}$, where $\alpha_{p}$ is
the largest index of $p$ dividing $n$.

Finally, the solutions
of~\eqref{eqn-coprime-solns-2} modulo $q^{\alpha_{q}
}$ for every
prime $q|n$ can be combined using Chinese Remaindering to get
a
solution modulo $n$.
\end{proof}

It is natural to try to obtain bounds on the diameter of $\cG=G(n;S)$
in terms of $\# D$. Certainly we have trivial bounds
$$
2 \le \diam \cG \le 2 \# D + 1.
$$
The following result shows that in general no better bounds are possible.

\begin{theorem}
\label{thm-diam-bounds-tight}The following
statements are true for integral
circulant
graphs:

\begin{enumerate}
\item[i] For $r\geq3$, let $n$ be the
product of distinct odd primes
$p_{1},\ldots,p_{r}$ and let
$D=\{p_{1},\ldots,p_{r}\}$. The graph
corresponding to these
parameters has diameter $2$.

\item[ii] Let $m$ be the product of
distinct odd primes
$p_{1},\ldots,p_{r}$. Let
$n=2m^{2}$
and
\[
D=\left\{(m/p_1)^2 , \ldots, (m/p_r)^2\right\}.
\]
The graph
corresponding to these parameters has diameter
$(2r+1)$.
\end{enumerate}
\end{theorem}

\begin{proof} \emph{Part~i}.
By the hypothesis $n=p_{1}\ldots p_{r}$, $D=\{p_{1},\ldots
,p_{r}\}$. Recall that $T=\{0\}\cup_{d\in D}G_{n}(d)$. Let $\cG$ be
the corresponding graph. We show that given any $\ell\in\Z_n$, we
have $\ell\in T+T$.

Suppose $\ell$ is coprime to $n$. Then using the methods of
Theorem~\ref{thm-diam-bounds}, we can find a solution $x_{1},
x_{2}\in \Z_n^{\ast}$, such that $p_{1}x_{1}+p_{2}x_{2} \equiv \ell
   \pmod n$. Thus, $\ell\in T+T$. If $\ell$ is not coprime
to $n$ then without loss of generality, we can  assume that
$p_{1}|n$. Again, using the methods of Theorem~\ref{thm-diam-bounds}
we can find a solution $x_{1},\ x_{2}\in\Z_n^{\ast}$ such that
$p_{1}x_{1}+p_{1}x_{2} \equiv \ell \pmod n$. Thus, $\ell\in T+T$.

Therefore, $T+T=\Z_n$. As the smallest additive generator
set
contained in $D$ is of size $2$ we deduce
from
Theorem~\ref{thm-diam-bounds} that $\diam
\cG=2$.

\emph{Part~ii}.  We recall that $T = \{0\}\cup_{d\in D}G_{n}(d)$.
Let $\cG$ be the corresponding graph. We  show that $m\not \in 2rT$.

Suppose that $m\in 2rT$. This means that there are

$d_{1},\ldots,d_{2r} \in D$ such that $m\in
G_{n}(d_{1})+\ldots+G_{n}(d_{2r})$. Since $p_{j}^{2}\nmid  m$, we
deduce that $(m/p_j)^2\in\{d_{1},\ldots,d_{2r}\}$, $j=1, \ldots, r$.

Without loss of generality, we can  assume that
$d_{1}=(m/p_1)^2,\ldots, d_{r}=(m/p_r)^2$. In other words, there are
$x_{1},\ldots,x_{2r}\in \Z_n^{\ast}$ such that
\begin{equation}
\frac{m^{2}}{p_{1}^{2}}x_{1}+\ldots+\frac{m^{2}}{p_{r}^{2}}x_{r}
+d_{r+1}x_{r+1}+\ldots+d_{2r}x_{2r}
\equiv   m  \pmod
n.
\label{eqn-diam-bounds-tight-1}
\end{equation}
Taking  the above congruence modulo $p_{1}$ we deduce that
\[
\frac{m^{2}}{p_{1}^{2}}x_{1}+d_{r+1}x_{r+1}+\ldots+d_{2r}x_{2r}
\equiv 0 \pmod {p_1}. \] As $\gcd(x_{1}, p_{1}) =1$, the above
congruence implies $(m/p_1)^2 \in \{d_{r+1},\ldots,d_{2r}\}$.
Similarly, taking~\eqref{eqn-diam-bounds-tight-1} modulo primes
$p_{2},\ldots,p_{r}$ and repeating the argument we deduce

$$(m/p_1)^2,\ldots,(m/p_r)^2\in\{d_{r+1},\ldots,d_{2r}\}.$$ Without
loss of generality, we can assume that $$ d_{r+1}
=\frac{m^{2}}{p_{1}^{2}},\ldots, d_{2r}=\frac{m^{2}}{p_{r}^{2}}. $$
Thus, the congruence~\eqref{eqn-diam-bounds-tight-1} becomes $$
\frac{m^{2}}{p_{1}^{2}}(x_{1}+x_{r+1})+\ldots+\frac{m^{2}}{p_{r}^{2}}
(x_{r}+x_{2r}) \equiv m \pmod n.$$ Recall that $x_{1},\ldots,x_{2r}$
are coprime to $n$. So, looking at the above equation modulo $2$, we
deduce $m \equiv 0 \pmod 2$, which is a contradiction as $m$ is odd.

This shows that $m\not \in 2rT$ and hence $\diam \cG>2r$. Since the
smallest additive generator set of $\Z_n$ in $D$ is of size $r$, by
Theorem~\ref{thm-diam-bounds}, we have that $\diam \cG=2r+1$.
\end{proof}

\section{Conclusion}
\label{sec:concl}

We have proved that, a quantum system, whose hamiltonian is
identical to the adjacency matrix of a circulant graph, is periodic
if and only if the graph is integral.

We have bounded the number of vertices of integral circulant graphs
in terms of their degree, characterized bipartiteness and given
exact bounds for the diameter.

It is a natural problem to extend the theorems~\ref{thm:periodic},
\ref{thm:Nkb} and~\ref{thm-diam-bounds} to other classes of Cayley
graphs. For example, Cayley graphs of Abelian groups.

We conclude with a partial result about perfect state transfer. We
say that there is \emph{perfect state transfer} (see~\cite{CDEL}) in
a graph $\cG$ between the vertex $a$ and the vertex $b$ if there is
$0<t<\infty$, such that
\[
|\langle a|e^{-\iota A(\cG)t}|b\rangle|=1.
\]
For an integral circulant graph $\cG =G(n;S)$, we have the following
setting: for all $0\leq j\leq n-1$,
$v_{j}=[1,\omega^{j},\ldots,\omega^{j(n-1)}]^{T}$ is an eigenvector
of $A(\cG)$ corresponding to the eigenvalue $\lambda_j$
given by~\eqref{eq-lambda-in-terms-of-omega}.
Thus
$$A(\cG)=\frac{1}{n}\sum_{j=0}^{n-1}\lambda_{j}v_{j}v_{j}^{\dagger}.
$$
This gives
$$e^{-\iota
A(\cG)t}=\frac{1}{n}\sum_{j=0}^{n-1}e^{-i\lambda_{j}t}v_{j}
v_{j}^{\dagger}$$
and
$$|\langle a|e^{-i
A(\cG)t}|b\rangle|=\sum_{j=0}^{n-1}e^{-i\lambda_{j}t}\omega^{\ell(a-b)}.
$$

We have then the next question: are there $0\leq a,b\leq(n-1)$ and
$t\in\R$ such that $|\langle a|e^{-i A(\cG)t}|b\rangle|=1$?

\begin{proposition}
If $n$ is odd then there do not exist $0\leq a<b\leq(n-1)$ and
$t\in\R^{>0}$ such that
$\vert \left<a\vert e^{\iota A t}\vert b\right> \vert=1$. In other
words, an integral circulant
graph having odd number of vertices cannot have perfect state transfer.
\end{proposition}
\begin{proof}
Recall
$$e^{\iota At}=\frac{1}{n}\sum_{\ell=0}^{n-1}e^{\iota\lambda_\ell t} v_\ell
v_\ell^\dagger.$$ Therefore,
\begin{align*}
\left<a|e^{\iota At}|b\right> =&
\frac{1}{n}\sum_{\ell=0}^{n-1}e^{\iota\lambda_\ell t}
\omega^{\ell a}\omega^{-\ell b}\\
=&\frac{1}{n}\sum_{\ell=0}^{n-1}e^{\iota\lambda_\ell t}\omega^{\ell
(a- b)}.
\end{align*}
Now the magnitude of the above expression is clearly $\le 1$. The
equality holds if and only if each term is $1$ implying that
$e^{\iota\lambda_\ell t}=\pm 1$ and $\omega^{\ell(a- b)}=\pm 1$ for
all $\ell$. Now if $n$ is odd then $\omega^{\ell(a- b)}=\pm 1$
happens only when $a \equiv b\pmod n$. Thus, there is no perfect
state transfer when $n$ is odd.
\end{proof}

When $n$ is even there is  perfect state transfer (between
vertices $a$ and $a+\frac{n}{2}$) if there exists a $t\in\R^{>0}$
such that $e^{\iota\lambda_\ell t}=(-1)^{\ell}$ for all
$\ell\in\{0,\ldots,n-1\}$. For example, this happens in the case of
$n=4$ and $S=\{1,3\}$. However, we do not know whether there are
other such instances.

\section*{Acknowledgments}
Part of this work has been carried out while the second author was
visiting CWI. This was possible thanks to the financial support of
CWI and the kind hospitality of Harry Buhrman.

\end{document}